\documentclass[aps,pre,twocolumn,showpacs]{revtex4-1}
\usepackage[dvips,pdftex]{graphicx}
\usepackage{hyperref}
\usepackage{amsmath}
\usepackage{amssymb}
\usepackage[figuresright]{rotating}
\usepackage{braket}
\usepackage{bm}
\usepackage{amsmath,amssymb}
\usepackage{color}
\usepackage{url}
\usepackage{algorithm}
\usepackage[titletoc]{appendix}
\usepackage{bbm}
\usepackage{bbold}
\usepackage{mathrsfs}
\usepackage[capitalize]{cleveref}

\usepackage[utf8]{inputenc}
\newcommand{\id}{\mathbb{1}}
\newcommand{\zb}{\bar{z}}
\newcommand{\wb}{\bar{w}}
\newcommand{\cG}{{\cal G}}
\newcommand{\cR}{{\cal R}}
\newcommand{\cB}{{\cal B}}
\newcommand{\cX}{{\cal X}}
\newcommand{\cL}{{\cal L}}
\newcommand{\btr}{\textup{bTr}}
\newcommand{\tr}{\textup{Tr}}

\newcommand{\<}{\left<}
\renewcommand{\>}{\right>}
\def\oper{{\mathchoice{\rm 1\mskip-4mu l}{\rm 1\mskip-4mu l}
{\rm 1\mskip-4.5mu l}{\rm 1\mskip-5mu l}}}

\begin{document}
\title{Universal spectra of random Lindblad operators}

\author{Sergey Denisov$^{1}$, Tetyana Laptyeva$^2$, Wojciech Tarnowski$^3$,
Dariusz Chru{\'s}ci{\'n}ski$^4$  and
Karol \.{Z}yczkowski$^{3,6}$}

\affiliation{$^1$ Department of Computer Science, Oslo Metropolitan University, N-0130 Oslo, Norway}
\affiliation{$^2$Department of Control Theory and Systems Dynamics, Lobachevsky University, Gagarina Av.\ 23, Nizhny Novgorod, 603950, Russia}
\affiliation{$^3$Marian Smoluchowski Institute of Physics,
    Uniwersytet Jagiello\'{n}ski, Krakow, Poland}
\affiliation{$^4$Institute of Physics, Faculty of Physics, Astronomy
and Informatics \\  Nicolaus Copernicus University, Grudzi{a}dzka
5/7, 87--100 Torun, Poland}
\affiliation{$^5$Centrum Fizyki Teoretycznej PAN, Warszawa, Poland}

\date{May 13, 2019
}

\begin{abstract}
 To understand typical dynamics of an open quantum system in continuous time, we introduce
 an ensemble  of random Lindblad operators, which generate
Markovian completely positive  evolution in the space of density matrices. 
Spectral properties of these operators,
including the shape
of the spectrum in the complex plane,
are evaluated by using methods of free probabilities and explained with non-Hermitian random matrix models. We also demonstrate  universality of the spectral features. The
notion of ensemble of random generators of Markovian qauntum evolution constitutes a step towards categorization of dissipative quantum chaos.
\end{abstract}

\maketitle

\noindent \textit{Introduction.}  Any real system is never perfectly isolated from its environment and the theory of open quantum systems \cite{open1,open2,open3} provides appropriate tools to deal with such phenomena as quantum dissipation and  decoherence.
In the Markovian regime (which assumes a weak interaction between the system and its environment and separation of system and environmental time scales), the evolution of an $N$-level open quantum system can be modeled by using the  master equation $\dot{\rho}_t = \mathcal{L}(\rho_t)$.
The corresponding Markovian generator $\mathcal{L}$ (often called a Lindblad operator or simply {\em Lindbladian} \cite{open1,Alicki}) has the well known Gorini-Kossakowski-Sudarshan-Lindblad form (GKSL),
\begin{eqnarray}  \label{L0}
\mathcal{L}(\rho) = -i[H,\rho]+\mathcal{L}_D(\rho) = \mathcal{L}_U(\rho) +\mathcal{L}_D(\rho),~~
\label{eq:0}
\end{eqnarray}
with the dissipative part
\begin{eqnarray}  \label{L1}
\!\!\! \mathcal{L}_D(\rho) =
\!\! \!\sum\limits_{m,n=1}^{N^2-1} \!\! K_{mn} [F_n\rho F^{\dagger}_m - \frac 12 (F^{\dagger}_m F_n\rho+\rho F^{\dagger}_m F_n)],~~
\label{eq:1}
\end{eqnarray}
where traceless matrices  $\{F_n\}$, $n = 1,2,3,\ldots,N^2-1$, satisfy
orthonormality condition, $\mathrm{Tr}(F_n F_m^\dagger) = \delta_{n,m}$. Finally, the
complex {\em Kossakowski matrix}
 $K = \{K_{mn}\}$  is positive semi-definite.  The solution of the master equation $\dot{\rho}_t = \mathcal{L}(\rho_t)$ gives rise
 to the celebrated Markovian semigroup $\Lambda_t = e^{t\mathcal{L}}$, such that for any $t \geq 0$ map $\Lambda_t$ represents a quantum
 channel, a completely positive and trace-preserving linear map \cite{GKSLhistory}.

In this Letter we analyze spectral properties of random Lindblad operators. Spectral analysis lies in the heart of quantum physics. In the static case, the spectrum of the Hamiltonian provides the full information about possible states of the system and system evolution.
Such analysis plays also a key role in the study of dissipative quantum evolution -- eigenvalues and eigenvectors of the Lindblad operator provide the full information about the dynamical properties of the open system \cite{zanardi17}.  
Spectra of dynamical maps were recently addressed
in Ref.~\cite{CMM} in connection to quantum non-Markovian evolution. This connection was experimentally verified recently  \cite{spectra-PRL}, which proves that spectral techniques can be used to characterize non-Markovian behavior as well. Here, instead of analyzing specific physical models (like in Refs.~\cite{CMM,spectra-PRL}), we look for universal spectral properties displayed by generic Lindblad operators. It should be stressed that the standard examples of generators of order $N=2$,
usually considered in the literature,
do not display universal features.
We address the problem when $N >> 1$ by using the powerful apparatus of Random Matrix Theory (RMT) \cite{Mehta}.

RMT already found many  applications in physics. It started form  the Wigner statistical
approach to nuclear physics, with his celebrated surmise for the distribution of energy level spacings in complex nuclei~\cite{WignerSurmise},   and a series of Dyson papers on statistical theory of spectra \cite{Dyson, DysonMehta,DysonBrownian,DysonThreefold}. It was soon realized that quantum dynamics corresponding to classically chaotic dynamics can be described by suitable ensembles of random matrices \cite{Haake,qchaos,brown}. In the case of autonomous
quantum systems, one could mimic Hamiltonians with the help of ensembles of random Hermitian
matrices invariant with respect to certain transformations. Depending
on the symmetry properties of the system investigated, one may use orthogonal,
unitary, or symplectic ensembles \cite{Mehta}. In the case of time-dependent, periodically
driven systems, corresponding unitary evolution operators can be described by one of three circular ensembles of Dyson \cite{DysonThreefold}.

\begin{figure*}[t]
	\includegraphics[angle=0,width=0.95\textwidth]{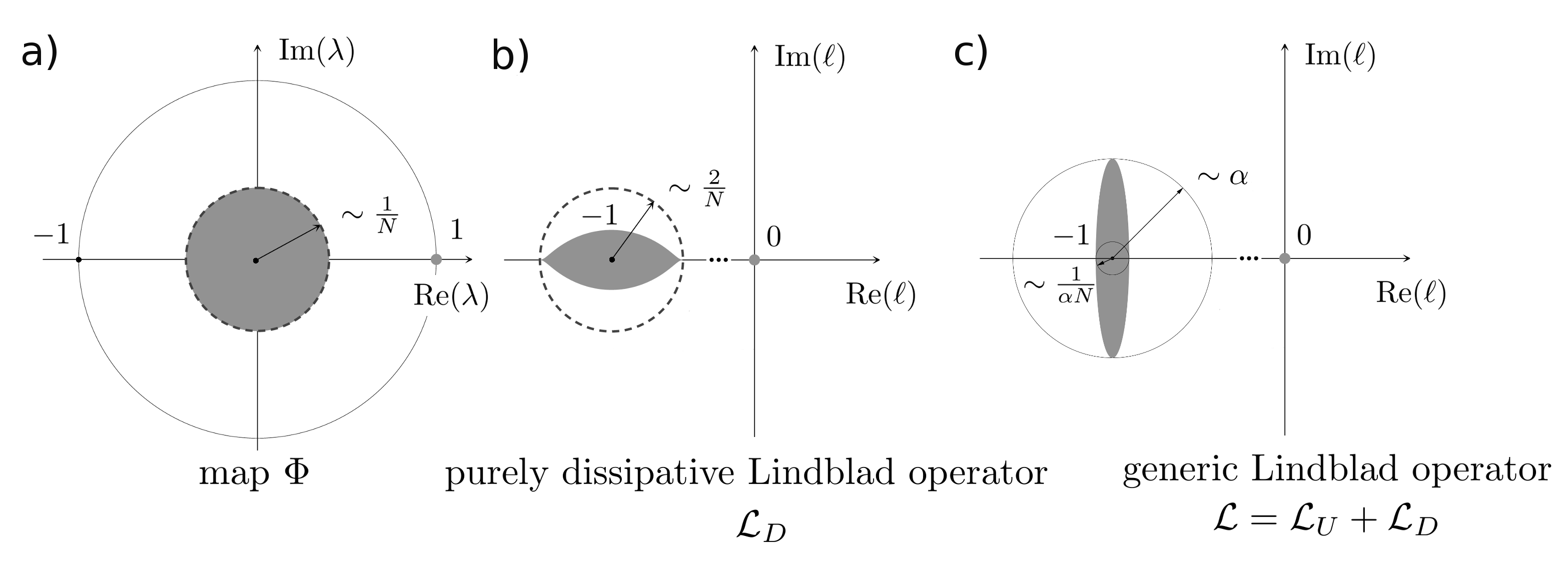}
	\caption{Eigenvalue localization areas (grey) 
	of (a) a random quantum channel, (b) random purely  dissipative  Lindblad operator, Eq.~(3), and (c) random generic Lindbladian with relative weight of the unitary component $\alpha$, Eq.~(8). 	While for the random channel (a) the distribution approaches a  Girko's disc upon the increase of the number of system levels $N$ \cite{BCSZ09}, eigenvalues of the random purely dissiaptive Lindblad operator (b) fill the interior of the universal lemon-like contour, Eq.~(7). For the Lindblad operator with the unitary component (c), the spectral boundary approaches an ellipse  upon increase of $\alpha$.}
	\label{fig:1}
\end{figure*}

Similar ideas found applications in disorder systems, single-particle \cite{Mirlin} and many-body \cite{Serbyn} ones. From a different perspective, a deep connection to RMT was observed in the models of 2D quantum gravity \cite{QG,QG2} and gauge theories with  the large gauge group $U(N)$ \cite{Parisi}.

In the case of discrete dynamics described by quantum operations,
various ensembles of random channels are known \cite{BZ17},
including the ensemble in which maps are generated from the
entire convex set of quantum operations according to the flat,
Hilbert--Schmidt measure \cite{BCSZ09}. Recently, powerful methods of random matrices found interesting applications in quantum information theory \cite{CN16,AS17}. For instance, techniques  based on random
operations were used by Hastings to refute the celebrated additivity conjecture
concerning the minimal output entropy of quantum channels \cite{Ha09}.

In the case of continuous quantum dynamics, a class of
random Lindblad equations with decay rates
obtained by  tracing out a random reservoir,
was studied  in Refs. \cite{Bu05,BG09}. RMT was applied to open quantum systems in the context of scattering matrices and non-Hermitian effective Hamiltonians, see \cite{open-random} for a review.

Our perspective in this paper is entirely  different: We introduce an ensemble of random  Lindblad
operators, which describe generic continuous time evolution of
an  $N$-level open quantum system, and evaluate universal properties of operator spectra.
Namely,  we analyze the distribution of eigenvalues of a randomly chosen operator $\mathcal{L}$ and study the scaling of spectral characteristics  with  $N$. 


We start the analysis of random Lindbladians by briefly recalling main results concerning random quantum channels \cite{BCSZ09,BSCSZ10}.
Next we analyze the extreme case of purely dissipative evolution,
$H=0$ and  $\mathcal{L}=\mathcal{L}_D$. In this limit we can applies RMT, capture all essential features of the general scenario, and derive equation for the spectral boundary. Finally we address general case, when both unitary and dissipative components, $\mathcal{L}_U$ and $\mathcal{L}_D$, of the evolution generator, Eq.~(1), are present.

\begin{figure*}[t]
	\includegraphics[angle=0,width=1.\textwidth]{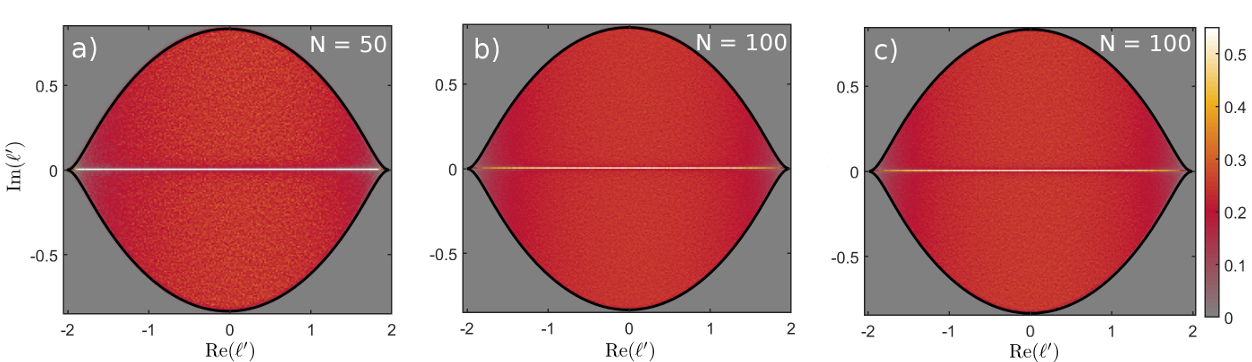}
	\caption{Spectral density 
	$P[\mathrm{Re}(\ell'),\mathrm{Im}(\ell')]$
	of the rescaled eigenvalues,	$\ell'= N (\ell +1)$,
	from  the spectrum
	of random purely dissipative Lindblad operators  ${\cal L_D}$  for  $N=50$ and $100$. We use two different sampling procedure for $N=100$,
	sampling the Kossakowski matrix from the Wishart ensemble (left and middle plots) and an alternative procedure \cite{SM} (right plot). Note a perfect agreement with the
	asymptotic boundary of the spectral bulk,
	Eq.~(7) (thick black line), derived with the random matrix model  (\ref{Lmodel}).
	Observe also a concentration of eigenvalues along the real axis, accompanied by depletion nearby  -- compare to Fig.~\ref{fig:3}a -- which decreases with $N$. Each distribution was sampled with $10^3$ realizations.}
	\label{fig:2}
\end{figure*}

\vspace{.2cm}

\noindent \textit{Random quantum channels.}  An ensemble of random channels (i.e., completely positive transformations \cite{open1,open2,open3}) $\Phi : M_N(\mathbb{C}) \to M_N(\mathbb{C})$, can be defined by the flat, Hilbert--Schmidt measure in the space
of all quantum operations.
It is known that the probability density of the corresponding
superoperator $\Phi$ acting on density matrices of order $N$
consists of the leading eigenvalue, $\lambda_{\rm 1} =1$,
corresponding to the invariant state, while all remaining eigenvalues fill a disk of radius $R=1/N$ centered at zero; see Fig.~\ref{fig:1}a. {The bulk of the spectrum can be obtained by sampling random matrices $\frac 1N G_R$ \cite{BCSZ09} with $G_R$ being a member of a real Ginibre ensemble \cite{Gi65}. Recall that a real Ginibre ensemble of order $N$ is defined by i.i.d. matrix elements with ${\cal N}(0,1/N)$ and asymptotically the distribution of eigenvalues is uniform on the unit disk with a singular component at the real axis \cite{FN07,girko,tao}.}

Thus for a generic superoperator $\Phi$
the size of its spectral gap,  $\Delta_N=\lambda_1-|\lambda_2| =1-1/N$, increases with the
matrix dimension $N$, so the
convergence to equilibrium becomes exponentially fast.
For a large $N$ a typical channel becomes close to a one--step contraction,
which sends any state into the single invariant state,
$\Phi(\rho)=\rho_{\rm inv}=\Phi(\rho_{\rm inv})$.
It is known \cite{NPPB18} that a generic channel is close to be unital and
the correction term, $\Phi(\oper) - \oper$,
behaves like a random hermitian
matrix of the Gaussian unitary ensemble
with asymptotically vanishing norm.

\vspace{.4cm}

\noindent \textit{Purely dissipative random Lindblad operators.}  To generate a random operator $\mathcal{L}_D$, we fix an orthonormal Hilbert-Schmidt basis $\{F_n\}$ \cite{SM} and first sample a random Kossakowski matrix $K$. There many ways to do such sampling. However, as we show below, a particular way in which this
non-negative order $N^2-1$  matrix is sampled is not important: The spectral features of random purely dissipative Lindbaldians are {\it universal}.

The most natural way 
is to sample $K$ from the ensemble of square complex Wishart matrices,
distinguished by the fact that it induces
the Lebesgue measure in the space of quantum states \cite{ZPNC11}.
A Wishart matrix \cite{pastur} has the structure $W = GG^\dagger\ge 0$, where $G$ is a complex square Ginibre matrix with independent complex Gaussian entries. Such a choice
is physically motivated by the fact that these ensembles of random matrices
correspond to non-unitary evolution of quantum 
dynamical systems under the assumption of classical chaos \cite{Haake,BSCSZ10}.
%

We use the following normalization condition
$\mathrm{Tr}K = N$, that is, $K = N GG^\dagger/\mathrm{Tr}GG^\dagger$.
Note that eigenvalues of $K$, $\gamma_m$, $m = 1,...,N^2-1$,
which can be interpreted as decay rates \cite{open1},
are distributed according to the universal Marchenko-Pastur law \cite{pastur} with the mean value  $\<\gamma\>\sim 1/N$.
Diagonalizing the Kossakowski matrix one can reduce the form of $\mathcal{L}_D$ as follows:
\begin{equation}\label{VV}
  \mathcal{L}_D(\rho) = \sum_{m=1}^{N^2-1} \gamma_m [V_m \rho V_m^\dagger - \frac 12
  (V_m^\dagger V_m \rho + \rho V_m^\dagger V_m)] ,
\end{equation}
where $V_m$ are  called `noise' (or `jump') operators \cite{to_Campo}.
Thus $\Phi(\rho) = \sum_m \gamma_m V_m \rho V_m^\dagger$,
defines a Kraus representation of completely positive map. Moreover,  $\sum_m \gamma_m V_m^\dagger V_m = \Phi^\dagger(\oper)$, where $\Phi^\dagger$ is the dual map, ${\rm Tr}[A \cdot \Phi^\dagger(B)] =  {\rm Tr}[\Phi(A)\cdot B]$, and $\id$ is the identity matrix in $M_N(\mathbb{C})$. Therefore, Eq.~(\ref{VV}) can be rewritten as
\begin{equation}\label{L11}
  \mathcal{L}_D(\rho) = \Phi(\rho) -
  \frac 12 \bigl( \Phi^\dagger(\id) \rho + \rho\; \Phi^\dagger(\id) \bigr) ,
\end{equation}
which shows that the purely dissipative Lindblad generator is fully determined by a completely positive map $\Phi$.
If, in addition, $\Phi$ is trace-preserving, i.e., it is a quantum channel, we have $\mathcal{L}(\rho) = \Phi(\rho) - \rho$. This is not the case in general, and
Hermitian translation matrix,
$X=\Phi^\dagger(\oper) - \oper$,
does not vanish. Making use of this notation, we rewrite
the Lindblad operator as
${\cal L}_D(\rho) = [\Phi(\rho)-\rho] - \frac 12(X \rho + \rho X)$.

If $\Phi$ is a quantum channel, the spectrum of ${\cal L}_D$ is  the spectrum of $\Phi$ shifted by $-1$. Thus the leading eigenvalue, $\lambda_1 =1$, is translated into $\ell_1=0$ and the Girko disk
is now centered at $z=-1$.  Due to the trace preserving quantum Markovian dynamics, Lindblad generators have always a zero eigenvalue.

\begin{figure}[b]
\begin{center}
\includegraphics[width=0.4\textwidth]{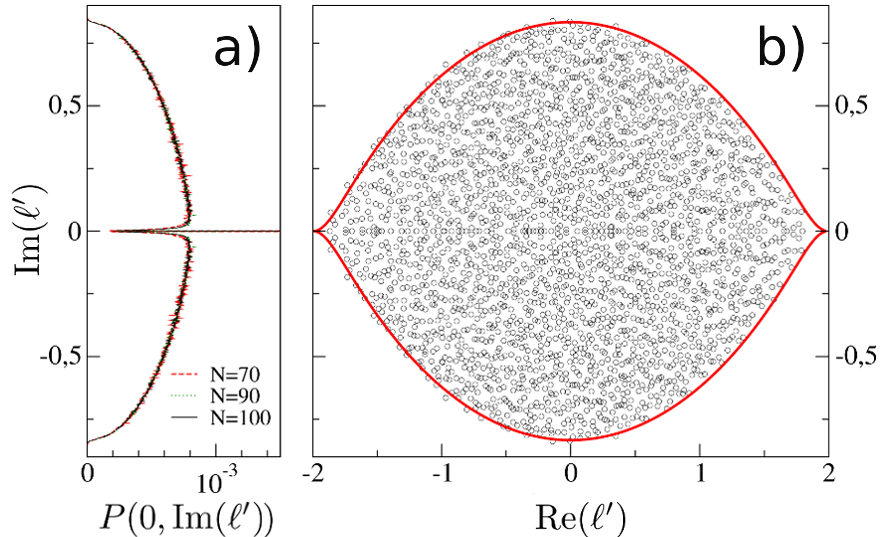}
\end{center}
\caption{(a) Marginal distribution, $\mathrm{Re}(\ell')=0$, of  rescaled eigenvalues for three different values of $N$. (b)
Rescaled eigenvalues $\ell'$ (empty dots) of a single  Lindblad operator realization for $N=50$. Red outer contour is the  boundary derived from the random matrix model, Eq.~(6).
\label{fig:3}
}
\end{figure}

To sample spectra of random Lindbladians, we generate $10^3$
realizations for different values of $N$, ranging from $30$ to $100$.
In order to reveal the universality of  spectra of the  operators,
it is useful to apply an affine transformation,
${\cal L}_D'= N ({\cal L}_D+1)$
\cite{SM}.
Then the bulk of the spectrum of ${\cal L}'$
becomes scale invariant and independent of $N$, see Figs.~\ref{fig:2}
and ~\ref{fig:3}(a).

\begin{figure*}
	\includegraphics[angle=0,width=1.\textwidth]{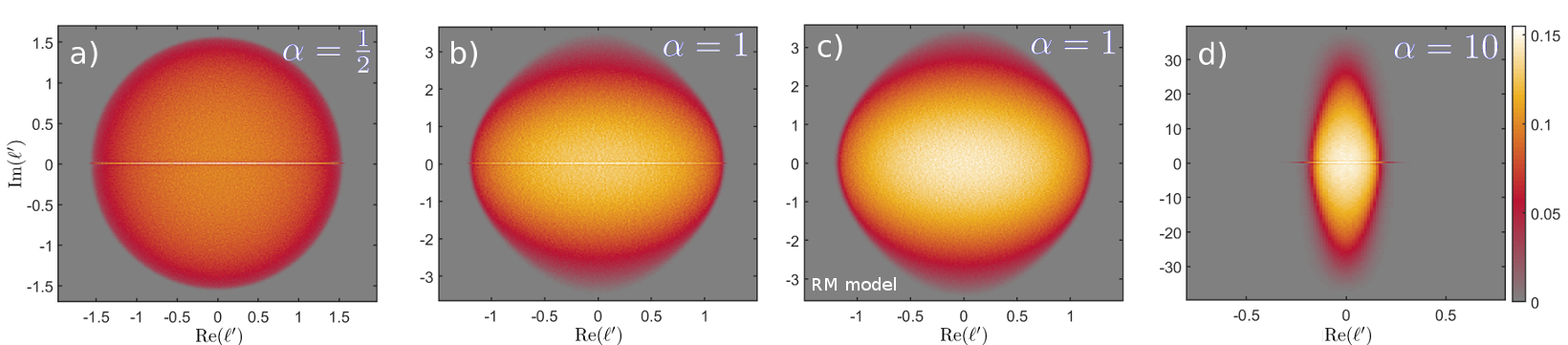}
	\caption{Probability distributions
	$P[\mathrm{Re}(\ell'),\mathrm{Im}(\ell')]$
	of the rescaled eigenvalues,	$\ell'= N (\ell +1)$,
	from the spectrum
	of random Lindblad operators  ${\cal L}$, Eq.~(8), for $N=100$ and different values of the unitary component weight $\alpha$. (a,b,d): We present here the results of the  sampling of the Kossakowski matrix from the Wishart ensemble, emphasizing that alternative generation procedures \cite{SM} yield the same results.  Panel (c) presents spectral distribution obtained with the random matrix model, Eq.~ (10). Each distribution was sampled with $10^3$  realizations. Additional normalization of the densities is performed in order to keep maximal values of all distributions  equal.}
	\label{fig:5}
\end{figure*}

The spectral density inside the 'lemon' is manifestly non-uniform.
Also notable is the eigenvalue concentration along the real axis and the corresponding  depletion near by, see Fig~\ref{fig:3}a.
Although ${\cal L}_D$ is represented by a complex matrix,
it can be made real by a similarity transformation,
which explains the effect of concentration \cite{FN07,tao,Edelman}.

The shape of the sampled eigenvalue distribution $P[\mathrm{Re}(\ell'),\mathrm{Im}(\ell')]$ is significantly
different from the Girko disk
and  displays a universal lemon–like shape.
It is  noteworthy that  already for $N  = 50$, a  {\em single} realization is  enough to reproduce the universal shape, see Fig.~\ref{fig:3}b. From the scale invariance  it follows that the spectral gap of  ${\cal L}_D$ scales as $\Delta_N \simeq 1-\frac 2N$.
It is clear that the very term $(X \rho + \rho X)$ is responsible for the  `disk $\to$ lemon' deformation. 

Finally, we performed sampling by using alternative generation procedures \cite{SM} and for $N \geq 50$ obtained near identical results (differences are within the sampling errors); see Fig. 2 (c). This confirms the universality of the spectral distribution.

\vspace{.2cm}

\noindent \textit{Random matrix model} --- Here we consider a  RMT model explaining the observed  spectral properties of purely dissipative Lindblad generators.  Let us recall that the spectrum of $\mathcal{L}_D$ represented in (\ref{L11}) coincides with the spectrum of the following $N^2 \times N^2$ complex matrix $\mathcal{L}_{mn} = {\rm Tr}[F_m \mathcal{L}_D(F_n)]$. This matrix becomes real if basis matrices $F_m$ are hermitian, due to the fact that $\mathcal{L}_D$ is hermiticity preserving. Another well known matrix representation of
the Lindblad operator reproducing its spectrum reads
\begin{equation}
\label{LPHI}
\widehat{\mathcal{L}}_D =
\widehat{\Phi} - \oper \otimes \oper -  \frac 12 ( X \otimes \oper + \oper \otimes {\overline X}) ,
\end{equation}
where $\widehat{\Phi} = \sum_{m=1}^{N^2-1} \gamma_m V_m \otimes \overline{V}_m$, and $\overline{V}_m$ stands for the complex conjugation. Note that $\widehat{\Phi}$ is neither hermitian nor real, however, the term $ X \otimes \oper + \oper \otimes {\overline X}$ is perfectly hermitian. To understand the
spectrum of $\mathcal{L}'_D$, we  use  matrix representation (5)
approximate its rescaled version with  the following random matrix model
\begin{equation}
\label{Lmodel}
\widehat{\mathcal{L}}_D' \approx \,
G_R 
-  (C \otimes \oper+ \oper \otimes 
  C).
\end{equation}
The matrix $G_R$ of size $N^2$ is taken from the real Ginibre ensemble.
The   correction term $C$  approximates $X$
by a symmetric GOE matrix \cite{NPPB18}
of size $N$.

Matrices are normalized as ${\rm Tr}G_RG_R^{\dagger}=N^2$,
so that its spectrum covers uniformly a  disk of radius $1$,
while ${\rm Tr}C^2=N/4$ assures that
its density forms the Wigner semicircle
of radius $1$.
The scaling and parameters of the model follows from the normalization of the Kossakowski matrix \cite{SM}.

We approach  spectral properties of the random matrix model~\eqref{Lmodel} with the quaternionic extension of free probability~\cite{JanikNowakFRV,JanikNowak2,FeinbergZee,FeinbergZee2,JaroszNowak}. Within this framework, we determine the  border of the spectrum of ${\cal L}_D'$ as given by the solution of the following equation involving a complex variable $z$ \cite{SM},
\begin{eqnarray}   \label{zGz}
\textup{Im}[z+G(z)]=0,
\end{eqnarray}
with
\begin{equation}
\label{G-quantum}
G(z) =  2z-\frac{2z}{3\pi}\left[(4+z^2)E\left(\frac{4}{z^2}\right)+(4-z^2)K\left(\frac{4}{z^2}\right)\right] \nonumber,
\end{equation}
where $E(k)$ and $K(k)$ are complete elliptic integrals of the first and second kind, respectively. The results of the sampling are in perfect agreement with this border, see Figs.~\ref{fig:2}-\ref{fig:3}. Evaluation of the spectral density inside the  'lemon' is much harder task; it could potentially be performed with diagrammatic techniques \cite{JanikNowak2}.

\vspace{.2cm}
\noindent \textit{General case of random Lindbladians.} 
Finally, we include unitary component $\mathcal{L}_U$ into  random Lindblad operator $\mathcal{L}$. For that we need random Hamiltonian $H$, which we sample  from the Gaussian Unitary Ensemble (GUE). 
To compare the spectrum of the general Lindblad operator with a purely dissipative one, we normalize the Hamiltonian, 
$\rm{Tr} H^2 = 1/N$, 
 and introduce a parameter $\alpha \geq 0$ which weights  contribution of the unitary component. The corresponding Lindbladians can be written as [see Eq.~(\ref{L11})]
\begin{equation}\label{L_ham}
  \mathcal{L}(\rho) =  -\frac{i\alpha}{\hbar}(H\rho - \rho H)+  \Phi(\rho) -
  \frac 12 \bigl( \Phi^\dagger(\id) \rho + \rho\; \Phi^\dagger(\id) \bigr).
\end{equation}
The sample spectra of the operator ${\cal L}'= N ({\cal L}+1)$,
for different values of $\alpha$, are shown on Fig.~4(a,b,d) (see also in \cite{SM}). Similar to the previously considered case of purely dissipative evolution, we find (as expected) a perfect scale invariance of the ${\cal L}'$ spectra starting from $N \geq 30$.

The shape of the spectra could be captured with the RMT. 
First, we transform  expression (\ref{L_ham}) into
\begin{equation}
\label{LPHI-1}
\widehat{\mathcal{L}} =
\widehat{\Phi} - \oper \otimes \oper -  \left(\frac 12  X + i\alpha H\right) \otimes \oper + \oper \otimes \left(\frac 12 {\overline X} - i\alpha \overline{H}\right).
\end{equation}
The spectrum of $\widehat{\mathcal{L}}$ can be explained by updating the  matrix model (\ref{Lmodel}),
\begin{equation}
\label{Lmodel-1}
\widehat{\mathcal{L}}' \approx \,
G_R 
-  (W \otimes \oper+ \oper \otimes 
  \overline{W}),
\end{equation}
where  $G_R$ of size $N^2$ is again taken from the real Ginibre 
ensemble, while the extended correction term $W = C + i \alpha H'$ 
contains now a random
GOE matrix $C$ and an anti-hermitian term proportional to 
a GUE matrix $H'$ of order $N$
normalized as ${\rm Tr}H'^2=N$.
Spectral density of the RMT model for $\alpha=1$ is shown in Fig.~4(c). It reproduced the density of the corresponding Linbladian ensemble (except of eigenvalue concentration at the real axis \cite{FN07,tao,Edelman}).

The eigenvalues 
of $W$ uniformly cover an ellipse with semi-axes  $\frac{1}{\sqrt{1+4\alpha^2}}$  and $\frac{4\alpha^2}{\sqrt{1+4\alpha^2}}$.
Spectral density of $\widehat{\mathcal{L}}'$ is therefore a (classical) convolution of two uniform densities supported on these ellipses followed by free convolution with the Girko disk
of unit radius; see sketch on Fig.~1(c). Contrary to the  case of purely dissipative Lindbladians, it is 
hardly possible to determine analytically spectral boundary of general Lindbladian ensembles. However, when $\alpha=\frac{1}{2}$, it immediately follows (since a convolution of two discs is a disc) that the spectral boundary is a circle.  This is in full agreement with the results of the sampling; see Fig. 4(a).


\vspace{.3cm}

\noindent \textit{Conclusions.} 
Our results constitute a step toward a spectral theory of dissipative Quantum Chaos \cite{brown,BSCSZ10}. Universal spectral features of different ensembles of unitary evolution generators -- that are Hamiltonians -- are the main pillar of the existing Quantum Chaos (QC) theory \cite{Haake}.  A notion of an ensemble of random operators of quantum Markovian evolution is therefore necessary
to extend QC into the area of open quantum systems. The two next steps would be (i) establishing  of links between the idea of 'typical Lindbladian' and physical models displaying chaotic dynamics (open systems that exhibit many-body localization at the Hamiltonian limit are prospective candidates \cite{mbl1,mbl2}), 
and (ii) evaluation of spectral properties of steady states of random Lindbladians. 

Finally, it should also be stressed that our approach works equally well in the classical case, where continuous dynamics in the space of  probability distributions is determined by  Kolmogorov generators \cite{else}.

\medskip
Note added:  One of the authors  (W.T.) attended a talk by Tankut Can given at
the conference in Yad Hashmona (Israel) in October 2018,
in which a parallel project on random Lindblad operators was presented. Since the first version of this work was posted 
in the arXiv in November 2018, three
other papers on the related subjects appeared  \cite{C19, COOG19,SRP19}.

\begin{acknowledgments}
{\sl Acknowledgements} S.D., D.C. and K.\.{Z}. appreciate the hospitality of the Center for Theoretical Physics of Complex Systems (IBS, South Korea) where this project was started.
D.C. and K.\.{Z}.  are grateful to {\L}. Pawela and Z. Pucha{\l}a for numerous discussions on random operations; they also acknowledge  support from Narodowe Centrum Nauki under
 the grant number 2018/30/A/ST2/00837 and   2015/18/A/ST2/00274, respectively.
S.D. and T.L. acknowledges  support by the Russian
Science Foundation via Grant No. 19-72-20086.
T.L. acknowledges support by the Basis Foundation (Grant No.18-1-3-66-1).
W.T. appreciates the financial support from the Polish Ministry of Science and Higher Education through ''Diamond Grant'' 0225/DIA/2015/44 and the
doctoral scholarship ETIUDA 2018/28/T/ST1/00470 from National Science Center.
\end{acknowledgments}

\vspace{4.ex}

\begin{widetext}

\makeatletter \renewcommand{\fnum@figure}{{\bf{\figurename~S\thefigure}}}
\setcounter{figure}{0}
\setcounter{equation}{0}
\renewcommand{\theequation}{S\arabic{equation}}

\section{Supplemental Material}

\subsection{Sampling of random Lindblad operators}

Due to the unitary equivalence, the particular choice of a
Hilbert-Schmidt basis to construct a Kossakowski matrix is not important. As $\{F_n\}$ ($n=1,\ldots,N^2-1$) we take the full set of infinitesimal generators of $SU(N)$ (see, e.g., \cite{Alicki}). Namely, let $|1\rangle,\ldots|N\rangle$ be an orthonormal basis in $N$-dimensional Hilbert space. The generators of $SU(N)$ are defined as the following $N^2-1$ Hermitian matrices:
\begin{itemize}
		\item $N(N-1)/2$ symmetric,
		
		$$S_{jk} = \frac{1}{\sqrt{2}}\Big(|j\rangle \langle k| + |k\rangle \langle j| \Big)\ , \ \ 1 \leq j < k \leq N $$
		
		\item $N(N-1)/2$  antisymmetric, 
		
		$$J_{jk} = -\frac{i}{\sqrt{2}} \Big(|j\rangle \langle k| - |k\rangle \langle j| \Big)\ , \ \ 1 \leq j < k \leq N$$
		
		\item and $N-1$ diagonal
		
		$$D_l = \frac{1}{\sqrt{l(l+1)}} \left( \sum_{k=1}^l |k\rangle \langle k|  - l |l+1\rangle \langle l+1| \right)\ , $$
		for $1 \leq l \leq N-1$.
\end{itemize}
For $N=2$ this recipe  yields Pauli matrices while for $N=3$ it results in the standard eight Gell-Mann matrices.

A brute-force sampling by using Eq.~(1) becomes extremely slow and ineffective already with $N=30$. A single-core implementation of such sampling cannot produce a single realization for $N=100$ on the time scale of several days.
To overcome this bottleneck, we parallelize the sampling procedure and realize it on a large computational cluster.
The detailed information will be presented in a separate paper [S1]; here we only briefly outline two key steps.

First, in order to multiple matrices in Eq. (1), we avoided standard built-on matrix-matrix multiplications because the
matrices to multiple a Kossakowski matrix with are very sparse. So this operation has been encoded explicitly, in the element-wise manner. 

Second, calculation of the summands on the rhs of Eq.~(1), together with corresponding multiplications,  was performed in parallel, on several cores simultaneously, and then the results were summed up.

Sampling simulations  were performed on the Lobachevsky supercomputer (Nizhny Novgorod) and the MPIPKS cluster  (Dresden).

\subsection{Sampling of the Kossakowski matrix\label{sec:K-sampling}}

Under normalization condition $\mathrm{Tr}K = N$, sampling of the positive semi-definite matrix $K$ reduces to the sampling of a random density matrix [SM2].
We considered different sampling procedures listed in Table I of Ref.~[SM2], 
\begin{eqnarray}
K = N\frac{SS^\dagger}{\mathrm{Tr}SS^\dagger},~~~~S[k,s] := [p_1U_1 + p_2U_2 +...+ p_kU_k]G_1G_2...G_s,
\end{eqnarray}
where $p=\{p_1,...,p_k\}$ is a random probability vector, $U_1,...,U_k$ 
is a set of  $k$ independent random unitary matrices distributed according to the Haar measure on $U(N)$ and $G_1,...,G_s$ is a set of  independent 
$N \times N$ random matrices sampled from the complex Ginibre ensemble. In the case $k=s=1$ it reduces to the sampling described in the main text (and which leads to the Marchenko-Pastur distributions 
of the $K$'s eigenvalues). We also used  combinations $\{k=1,s=2,3,7\}$ [leading to the Fuss–Catalan distributions $\pi(s)$] and $\{k=2,s=0,1\}$ [leading to the arcsine
and Bures ensembles [S2], respectively].
Finally, we used a more exotic sampling procedure,
\begin{eqnarray}
K = N\frac{UD}{\mathrm{Tr}UD}, 
\end{eqnarray}
with $U$ being a random unitary matrix sampled according to the Haar measure on $U(N)$ and
$D$ being a diagonal core of the singular-value decomposition (SVD), $G=VDW$,  of a random  matrix $G$ sampled from the complex Ginibre ensemble 
(the results of the sampling are shown in Fig. 2 (middle panel)  of the main text). In all cases we did not observe noticeable difference in the sampled spectral
densities (more formally, the differences were within sampling errors).

To summarize, it is not important, from the  spectral point of view, how the manifold of all random Lindlad generators, acting in the $N$-dimensional Hilbert-Schmidt 
space, is sampled -- provided that the sampling is not 'pathological' (f.e., $K$ is not restricted to a low-rank manifold) and normalization $\mathrm{Tr}K = N$ is kept.

In this section we also present distributions sampled for random Lindblad operators, with unitary component $\mathcal{L}_U$ included,  for $\alpha = 10$ and $\alpha = 100$; 
see Fig.~\ref{fig:GL}.

\begin{figure}[h]
    \centering
    \includegraphics[width=0.85\textwidth]{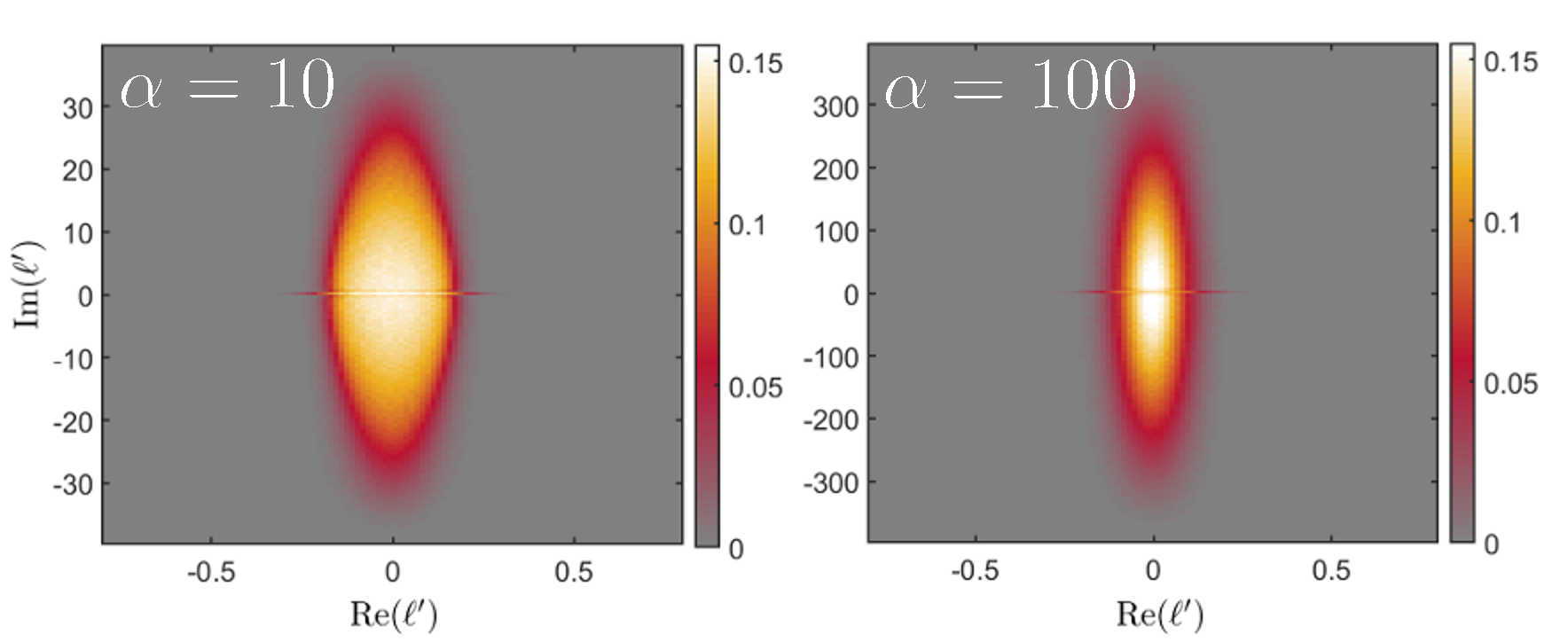}
    \caption{Probability distributions
	$P(\mathrm{Re}(\ell'),\mathrm{Im}(\ell'))$
	of the rescaled eigenvalues,	$\ell'= N (\ell +1)$,
	from the bulk of the spectrum
	of random Lindblad operators  ${\cal L}$, Eq~(8) in the main text, for  $N=100$ and two different values of
	the unitary component  weight $\alpha$. 
	We present here the results of the  sampling of the Kossakowski matrix from the Wishart ensemble emphasizing that alternative generation procedures yield the same results.  
	Each distribution was sampled with $10^3$  realizations. Additional normalization of the densities is performed in order to keep maximal values of all distributions  equal.
    \label{fig:GL}
    }
\end{figure}

\subsection{Justification of the random matrix  model\label{sec:Justification}
}

Since the Kossakowski matrix is positive definite, it can be decomposed as $K=Z^{\dagger}Z$. If the normalization condition $\tr K=N$ is relaxed to $\< \tr K\>=N$, then elements of $X$ 
are i.i.d. Gaussian random variables with the variance
\begin{eqnarray}
\<Z_{ab}Z^{\dagger}_{cd}\>=\frac{N}{(N^2-1)^2}\delta_{ad}\delta_{bc}. \label{eq:VarianceX}
\end{eqnarray}
We define a set of $N\times N$ matrices $Y_a=\sum_{m=1}^{N^2-1}Z_{am}F_m$ for $a=1,\ldots,N^2-1$. Note that due to independence of rows of $Z$, matrices $Y_a$ are independent. Moreover, 
probability distribution 
for elements of $Z$ is invariant under unitary transformations $Z\to ZU$, thus the statistical distribution of elements of $Y_a$ is the same, irrespective of the 
choice of basis matrices $F_m$. This convinces us that the entries of $Y_a$ are almost independent (the only constraint is $\tr Y_a=0$), thus in the large $N$ limit 
eigenvalues of $Y_a$ cover uniformly the disk of radius $r$, where
\begin{eqnarray}
r^2=\<\frac{1}{N}\tr Y_aY^{\dagger}_a\>=\frac{1}{N^2-1}.
\end{eqnarray}
Here we used the orthogonality of basis $\tr F_mF^{\dagger}_n=\delta_{mn}$ and \eqref{eq:VarianceX}.

Matrices $Y_a$ allow us to rewrite the Lindblad operator as
\begin{eqnarray}
\hat{\cL}=\hat{\Phi}-\oper\otimes \oper -(X\otimes \oper+\oper \otimes \overline{X}), \label{eq:LindbladMatrixRepr}
\end{eqnarray}
with
\begin{eqnarray}
\hat{\Phi}=\sum_{a=1}^{N^2-1}Y_a\otimes \overline{Y}_a, \\
2X=-\oper+\sum_{a=1}^{N^2-1}Y_a^{\dagger}Y_a.
\end{eqnarray}

All $N^2$ eigenvalues of $Y_a\otimes\overline{Y}_a$ are in the form of $\lambda_i\overline{\lambda}_j$ for $i,j=1,\ldots,N$, where $\lambda_{i,j}$ are eigenvalues of $Y_a$, 
thus their density is supported on a disk of radius $(N^2-1)^{-1}$.  $\hat{\Phi}$ is a sum of $N^2-1$ independent matrices $Y_a\otimes\overline{Y}_a$, thus, according to the 
central limit theorem for non-hermitian matrices, its spectral density is uniform on the disk of radius $(N^2-1)^{-1/2}$. As a consequence, in the large $N$ limit, $\hat{\Phi}$ 
can be modelled as a Ginibre matrix with the spectral radius $1/N$.

It is also clear that the matrix $W_a=\frac{1}{2}\left(Y^{\dagger}_aY_a-\frac{\oper}{N^2-1}\right)$ is a shifted and rescaled Wishart matrix, and its spectral density has zero 
mean and variance $\sigma^2=\frac{1}{4(N^2-1)^2}$. Since $X$ is a sum of $N^2-1$ independent such matrices, according to the central limit theorem for hermitian matrices, its 
spectral density is the Wigner semicircle supported on $[-1/N,1/N]$ with density
\begin{eqnarray}
\rho_X(x)=\frac{2N^2}{\pi}\sqrt{\frac{1}{N^2}-x^2}.
\end{eqnarray}
The above reasoning correctly predicts the $1/N$ scaling and unit shift $\hat{{\cal L}}=-\oper+\frac{1}{N}\hat{{\cal L}}'$ and justifies the following approximation of $\hat{{\cal L}}'$ 
\begin{eqnarray}\hat{{\cal L}}'\approx G_R-(C\otimes \oper+\oper \otimes \overline{C}). \label{eq:SMmodel}
\end{eqnarray}
While the matrix representation of Lindblad operator~\eqref{eq:LindbladMatrixRepr} is not real, $\overline{\hat{{\cal L}}}\neq \hat{{\cal L}}$, it is related to its complex 
conjugate via a similarity transformation $\overline{\hat{{\cal L}}}=P\hat{{\cal L}}P$ by a symmetric permutation matrix $P$ with a property that for any two matrices $A,B$ of size $N^2$
\begin{eqnarray}
P(A\otimes B)P=B\otimes A
\end{eqnarray}
holds. Therefore one can 
take $G_R$ as a real Ginibre matrix
and $C$ as a symmetric GOE 
matrix so that $\overline{C}=C$.

To take into account  Hamiltonian part of the dynamics our random matrix model~\eqref{eq:SMmodel}
  can be generalized to include also a purely antihermitian part,
\begin{equation}
 L= G_R - \oper \otimes \oper - \alpha (C \otimes \oper + \oper \otimes {\overline C})  +   i \beta (H \otimes \oper - \oper \otimes {\overline H}),
\end{equation}
which will be analyzed in a separate paper. Here $H$ denotes a Hermitian random matrix of order $N$ from the GUE  ensemble
  representing a generic Hamiltonian.

\subsection{A  spectrum of the matrix model for the Lindblad operator} \label{sec:SolvingMatrixModel}

\subsubsection{Quaternionic free probability}
Here we briefly review the quaternionic extension of free probability to nonhermitian random matrices, developed in [S3-S7]
(see also [S8]
for a recent rigorous treatment), focusing mostly on the aspects relevant for this study. For a  pedagogical introduction and  more explicit calculations we refer to [S9]

The main object of interest is the spectral density $\rho(z,\zb)=\left<\frac{1}{N}\sum_{i=1}^{N}\delta^{(2)}(z-\lambda_i)\right>$ on the complex plane. 
Here $\delta^{(2)}(z)=\delta(\textup{Re} z)\delta(\textup{Im} z)$. The density is obtained via the Poisson law $\rho(z,\zb)=\lim_{|w|\to 0}\frac{1}{\pi}\partial_{z\zb}\Phi(z,\zb,w,\wb)$, where $\Phi$ is 
the (regularized) electrostatic potential in two dimensions [S10],
\begin{equation}
    \Phi(z,\zb,w,\wb)=\<\frac{1}{N}\ln \det\left[(z-X)(\zb-X^{\dagger})+|w|^2\right]\>.
\end{equation}
To facilitate the calculations in large $N$ limit, one considers the generalized Green's function, which is a $2\times 2$ matrix
\begin{equation}
    \cG(Q)=\left<\frac{1}{N}\btr \left(Q\otimes \id-\cX\right)^{-1}\right>=\left(\begin{array}{cc}
    \cG_{11}     & \cG_{12} \\
    \cG_{21}     & \cG_{22}
    \end{array}\right),
\end{equation}
with 
\begin{eqnarray}
Q=\left(\begin{array}{cc}
z       & i\wb \\
    iw & \zb
\end{array} \right),\quad
\cX=\left(\begin{array}{cc}
    X & 0 \\
    0  & X^{\dagger}
\end{array}\right),\label{eq:QuatDef}
\end{eqnarray}
where we also introduced a block trace (partial trace) operation
\begin{eqnarray}
\btr\left(\begin{array}{cc}
A     & B \\
C     & D
\end{array}\right)=\left(\begin{array}{cc}
    \tr A & \tr B \\
    \tr C & \tr D
\end{array}\right).
\end{eqnarray}
Note that $Q$ is a matrix representation of a quaternion, thus we refer to this approach as quaternionic free probability. The upper-left element of $\cG$ yields spectral density via $\rho(z,\zb)=\lim_{|w|\to 0}\frac{1}{\pi}\partial_{\zb}\cG_{11}$, while the product of off-diagonal elements yields the correlation function
capturing non-orthogonality of eigenvectors [S11-S13].
An important fact for this paper is that the boundary of the spectrum can be determined from the condition $\cG_{12}\cG_{21}=0$.

Knowing the Green's function, one defines also the Blue's function as its functional inverse
\begin{eqnarray}
\cB(\cG(Q))=Q,\quad \cG(\cB(Q))=Q. \label{eq:GBrel}
\end{eqnarray}
Then, the quaternionic $R$-transform is defined as $\cR(Q)=\cB(Q)-Q^{-1}$, where the inverse is understood in the sense of $2\times 2$ matrix inversion. When two nonhermitian matrices $A$ and $B$ are free, then the $R$-transform of their sum is a sum of corresponding $R$-transforms
\begin{eqnarray}
\cR_{A+B}(Q)=\cR_A(Q)+\cR_B(Q). \label{eq:AdditionLaw}
\end{eqnarray}
In that sense, it generalizes the logarithm of the Fourier transform from classical probability to the noncommutative case.

\subsubsection{Nonhermitian Pastur equation}
We now consider a problem of finding a spectrum of the matrix $A+B$, where $A$ is a Ginibre matrix and $B$ can be arbitrary. Starting with \eqref{eq:AdditionLaw}, we add $Q^{-1}$ to both sides, obtaining
\begin{eqnarray}
\cB_{A+B}(Q)=\cR_A(Q)+\cB_B(Q).
\end{eqnarray}
Then we make a substitution $Q\to\cG_{A+B}(Q)$ and use the relation between Green's and Blue's function~\eqref{eq:GBrel}, obtaining
\begin{eqnarray}
Q-\cR_A(\cG_{A+B}(Q))=\cB_B(\cG_{A+B}(Q)).
\end{eqnarray}
In the next step we evaluate the Green's function of $B$ at both sides of equation and use~\eqref{eq:GBrel} to get
\begin{eqnarray}
\cG_B(Q-\cR_A(\cG_{A+B}(Q)))=\cG_{A+B}(Q), \label{eq:PasturEq}
\end{eqnarray}
which is the nonhermitian Pastur equation. In our case $A$ is Ginibre, the $R$-transform of which reads
\begin{eqnarray}
\cR_{A}(\cG_{A+B})=\left(\begin{array}{cc}
    0 & \cG_{12} \\
    \cG_{21} & 0
\end{array}\right),
\end{eqnarray}
thus \eqref{eq:PasturEq} simplifies to
\begin{eqnarray}
\cG_{B}\left[\left(\begin{array}{cc}
z     & -\cG_{12} \\
 -\cG_{21}     & \zb
\end{array}\right)\right]=\left(\begin{array}{cc}
\cG_{11}     & \cG_{12} \\
\cG_{21}     & \cG_{22}
\end{array}\right), \label{eq:PasturSimp}
\end{eqnarray}
where we suppressed index `$A+B$' when writing components of $\cG_{A+B}$.
We also used the fact that all important quantities are calculated in the $|w|\to 0$ limit and took this limit at the level of this algebraic equation.

\subsubsection{Embedding of hermitian matrices}
Equation \eqref{eq:PasturSimp} is true for general (even not necessarily random) matrix $B$, but the quaternionic Green's function can be easily obtained for Hermitian matrices. In such a case it  reads [S6]
\begin{equation}
    \cG(Q)=\gamma(q,\bar{q})\id_2-\gamma'(q,\bar{q})Q^{\dagger},
\end{equation}
with
\begin{eqnarray}
\gamma(q,\bar{q})=\frac{q G(q)-\bar{q}G(\bar{q})}{q-\bar{q}}, \\
\gamma'(q,\bar{q})=\frac{G(q)-G(\bar{q})}{q-\bar{q}},
\end{eqnarray}
where $q,\bar{q}$ are the eigenvalues of the $2\times 2$ quaternion matrix \eqref{eq:QuatDef} and $G(z)$ is the Stieltjes transform of the spectral density of $B$
\begin{eqnarray}
G(z)=\int_{-\infty}^{+\infty}\frac{\rho_B(x)dx}{z-x}.
\end{eqnarray}

\subsubsection{Border of the spectrum}
We are now ready to solve equation \eqref{eq:PasturSimp}. Focusing on the component $\cG_{12}$ of this matrix equation and using $\bar{\cG}_{21}=-\cG_{12}$, which follows from the definition of the quaternion~\eqref{eq:QuatDef}, we obtain
\begin{eqnarray}
-\frac{G(q)-G(\bar{q})}{q-\bar{q}}\cG_{12}=\cG_{12}.\label{eq:Bord1}
\end{eqnarray}
Now $q,\bar{q}$ are the eigenvalues of the matrix
\begin{eqnarray}
\left(\begin{array}{cc}
z     & -\cG_{12} \\
-\cG_{21}     & \zb
\end{array}\right). \label{eq:AuxMat}
\end{eqnarray}
There is one trivial solution, $\cG_{12}=0$, which is valid outside the spectrum. Inside the spectrum one has $\cG_{12}\neq 0$, but these two solutions match at the border of the spectrum, providing the equation for it. Note that for $\cG_{12}=0=\cG_{21}$ eigenvalues of \eqref{eq:AuxMat} are just $z,\zb$, thus we immediately obtain the equation for the borderline from \eqref{eq:Bord1}:
\begin{eqnarray}
G(z)-G(\zb)=\zb-z.
\end{eqnarray}
To solve a more general problem, namely spectrum of $A+\alpha B$ we use the fact that
the Stieltjes transform of the rescaled matrix $\alpha B$ is given by the Stieltjes transform of the original matrix $B$ via $G_{\alpha B}(z)=\frac{1}{\alpha}G_B(\frac{z}{\alpha})$, we get the final form:
\begin{eqnarray}
\textup{Im}\left[\alpha z+G\left(\frac{z}{\alpha}\right)\right]=0. \label{eq:BorderlineFinal}
\end{eqnarray}
In our model we set $\alpha=1$. To find the spectrum of the Kolmogorov generator, we take $B$ with Gaussian spectrum and then
\begin{eqnarray}
G_{\textup{class}}(z)=\frac{1}{\sqrt{2\pi}}\int_{-\infty}^{\infty} \frac{e^{-x^2/2}}{z-x}dx=\sqrt{\frac{\pi}{2}}e^{-z^2/2}\left(\textup{Erfi}\left(\frac{z}{\sqrt{2}}\right)-i \textup{sgn} (\textup{Im} z)\right)
\end{eqnarray}

\subsubsection{Stieltjes transform of $1\otimes C+C \otimes 1$}

In order to solve~\eqref{eq:BorderlineFinal}, we aim to find the Stieltjes transform of $B=1\otimes C+C \otimes 1$. Note that each eigenvalue of $B$ is of the form $\lambda=\mu_a+\mu_b$, 
where $\mu_{a,b}$ are the eigenvalues of $C$. Taking $C$ as GOE, the spectral density of which is the Wigner semicircle, $\rho_C(x)=\frac{2}{\pi}\sqrt{1-x^2}$, the spectrum of $B$ is 
therefore the (classical) convolution of two Wigner semicircles. This can be calculated using standard tools from probability. The Fourier transform of the Wigner 
semicircle is $\tilde{\rho}_C(k)=\frac{2}{k}J_{1}(k)$, where $J_1$ is the Bessel function of the first kind. Therefore the Fourier transform of $B$ reads $\tilde{\rho}_B(k)=\frac{4}{k^2}J_1^2(k)$.
To calculate the Stieltjes transform of $B$ we use the following representation $(z-x)^{-1}=\mp i\int_0^{\infty}e^{\pm ip(z-x)}dx$, which allows us to calculate the Stieltjes 
transform directly from the Fourier transform via $G(z)=\mp i\int_{0}^{\infty}e^{\pm ipz}\tilde{\rho}_B(\mp p)dp$, where we take the upper signs for $\textup{Im}z>0$ and lower 
for $\textup{Im}z<0$. The final result reads 
\begin{eqnarray}
G(z)=2z-\frac{2z}{3\pi}\left[(4+z^2)E\left(\frac{4}{z^2}\right)+(4-z^2)K\left(\frac{4}{z^2}\right)\right],
\end{eqnarray}
where $K(z)$ and $E(z)$ are the complete elliptic integrals of the first and second kind, respectively. Substitution into~\eqref{eq:BorderlineFinal} 
yields an implicit equation which is then solved numerically, see Fig.~\ref{fig:ModelSpectra} for comparison with the numerical simulations.

\begin{figure}[h]
    \centering
    \includegraphics[width=0.48\textwidth]{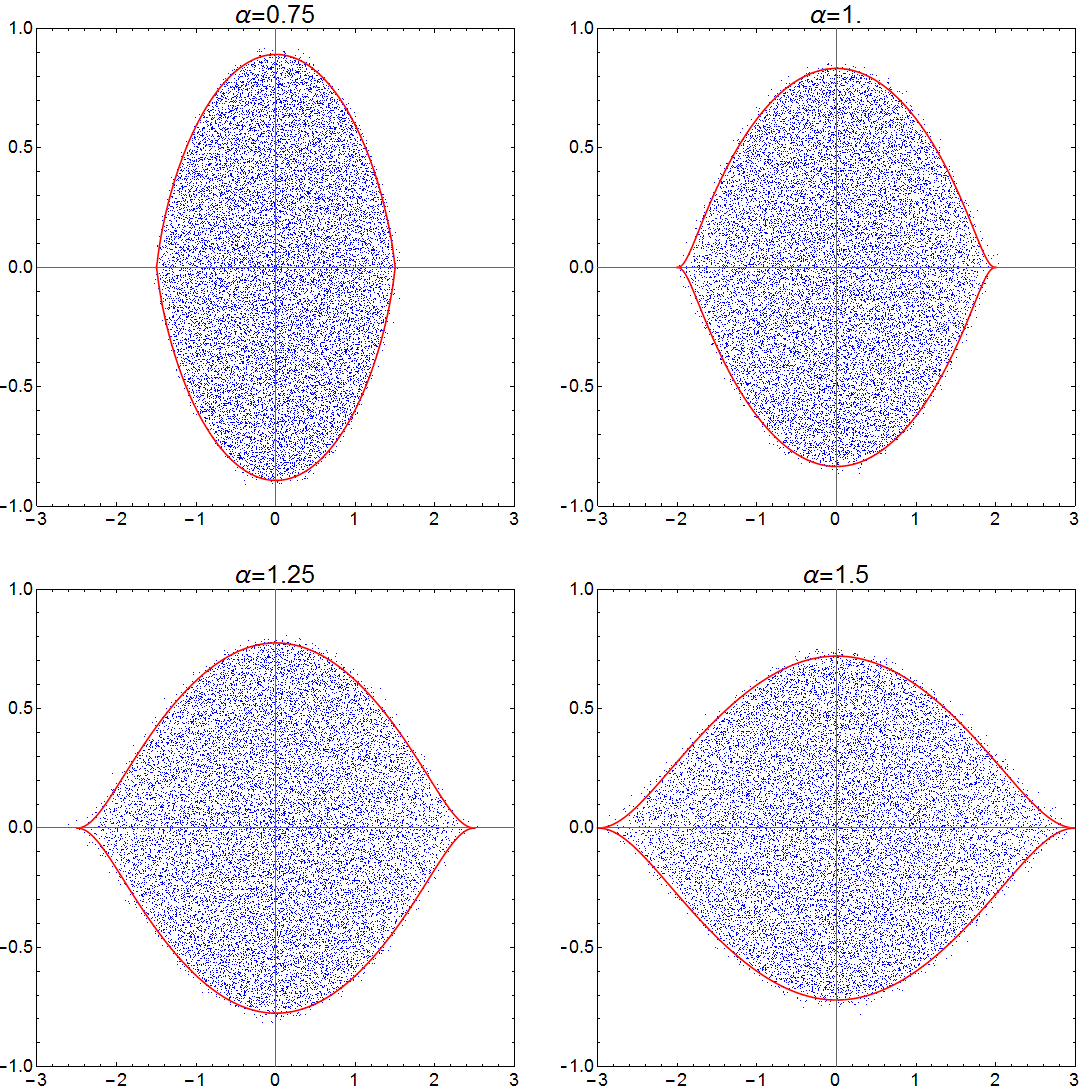}
    \caption{Spectra of the random matrix model $G_R+\alpha(C\otimes\oper+\oper\otimes C)$ for 
    $\alpha=0.75, 1.0, 1.25, 1.5$.
    The borderline of the asymptotic spectrum was calculated numerically from \eqref{eq:BorderlineFinal}. Points are obtained from the numerical diagonalization of 20 independent matrices of size $N^2=900$.
    \label{fig:ModelSpectra}
    }
\end{figure}

Interestingly, the technique used to derive the presented  results opens a new application area for the theory of free probability [S14]. 
Our usage of the theory is based on a two-step procedure: the classical convolution of two Hermitian
ensembles is followed by a free convolution with the Ginibre ensemble.

\vspace*{12pt}
\noindent\textbf{\large{{References}}}

\noindent [S1] I. Meyerov, A. Liniov, E. Kozinov, V. Volokitin, M. Ivanchenko, and S. Denisov, {\sl Unfolding quantum master equation into a system of linear equations:
computationally effective expansion over the basis of $SU(N )$ generators}, arXiv:1812.11626.

\noindent [S2]
K. {\.Z}yczkowski, K. A. Penson, I. Nechita, and B.  Collins,
{\sl Generating random density matrices}, 
J. Math. Phys. \textbf{E 52}, 062201 (2011).

\noindent [S3]
R. A. Janik, M. A. Nowak, G. Papp, J. Wambach and I. Zahed, {\sl 
Non-Hermitian random matrix models: Free random variable approach}, 
Phys. Rev. \textbf{E 55} (4), 4100 (1997).

\noindent [S4]
R. A. Janik, M. A. Nowak, G. Papp, I. Zahed, 
{\sl Non-hermitian random matrix models},
 Nucl.  Phys. \textbf{B 501} (3), 603 (1997).

\noindent [S5]
J. Feinberg, A. Zee, {\sl 
Non-Gaussian non-Hermitian random matrix theory: phase transition and addition formalism},
Nucl. Phys.\textbf{B 501} (3), 643 (1997).

\noindent [S6]
J. Feinberg, A. Zee, {\sl 
Non-hermitian random matrix theory: Method of hermitian reduction},
 Nucl. Phys. \textbf{B 504} (3), 579 (1997).

\noindent [S7]
A. Jarosz, M. A. Nowak, {\sl 
Random Hermitian versus random non-Hermitian operators—unexpected links},
 J. Phys. A: Math. Gen. \textbf{39} (32), 10107 (2006).

\noindent [S8]
S. T. Belinschi, P. \'{S}niady, R. Speicher, {\sl Eigenvalues of non-hermitian random matrices and Brown measure of non-normal operators: hermitian reduction and linearization method},
Linear Algebra Appl. \textbf{537}, 48 (2018).

\noindent [S9]
M.A. Nowak and W. Tarnowski, {\sl Spectra of large time-lagged correlation matrices from random matrix theory},
J. Stat. Mech.: Th. Exp. \textbf{2017}, 063405 (2017).

\noindent [S10]
H.J. Sommers, A. Crisanti, H. Sompolinsky and Y. Stein, {\sl 
Spectrum of large random asymmetric matrices},
Phys. Rev. Lett. \textbf{60},  1895 (1988).

\noindent [S11]
J. T. Chalker and B. Mehlig, {\sl Eigenvector statistics in non-Hermitian random matrix ensembles},
 Phys. Rev. Lett. \textbf{81}, 3367 (1998).

\noindent [S12]
B. Mehlig and J. T. Chalker, {\sl Statistical properties of eigenvectors in non-Hermitian Gaussian random matrix ensembles},
J. Math. Phys. \textbf{41}, 3233 (2000).

\noindent [S13]
R.A. Janik, W. N\"orenberg, M.A. Nowak, G. Papp, and I. Zahed, {\sl 
Correlations of eigenvectors for non-Hermitian random-matrix models},
 Phys. Rev.  \textbf{E 60}, 2699 (1999).

\noindent [S14] J. A. Mingo and R. Speicher, 
{\sl Free probability and random matrices}, (Springer Science,
New York, 2017).

\end{widetext}


\begin{thebibliography}{16}
\bibitem{to_Campo} Recently, Lindbladians with operators $V_m$  sampled from
Gaussian  Unitary Ensemble were considered in Ref.~\cite{Campo}.
This choice leads to dephasing-governed dynamics with the  uniform asymptotic state $\rho_{\rm inv} = \id/N$.
\end{thebibliography}
\end{document}